\documentclass[10pt,english,journal]{IEEEtran}
\usepackage[T1]{fontenc}
\usepackage[latin9]{inputenc}
\usepackage{geometry}
\usepackage{amsmath}
\usepackage{xpatch}
\usepackage{amssymb}
\usepackage{esint}
\usepackage{mathtools}
\usepackage{amsthm}
\usepackage{nicefrac}
\usepackage{bigints}
\usepackage{mathtools}
\usepackage{blkarray, bigstrut}
\usepackage{physics}
\usepackage{calligra}
\usepackage{graphicx}
\usepackage{epstopdf}
\usepackage{dsfont}
\usepackage{array,ragged2e}
\usepackage{enumitem}
\usepackage{etoolbox}
\usepackage{babel}
\usepackage[nopar]{lipsum}
\usepackage{psfrag}
\usepackage[ruled,lined,linesnumbered]{algorithm2e}
\usepackage{algorithmic}
\usepackage{algorithm2e}
\usepackage{balance}
\usepackage{float}
\usepackage{hyperref}

\usepackage{subcaption}

\SetKw{KwBy}{by}

\makeatletter
\newcommand{\removelatexerror}{\let\@latex@error\@gobble}
\makeatother

\graphicspath{ {Figures/} }

\xpatchcmd{\proof}{\hskip\labelsep}{\hskip5\labelsep}{}{}  
\makeatletter
\xpatchcmd{\proof}{\@addpunct{.}}{\@addpunct{:}}{}{}
\makeatother

\renewcommand\[{\begin{equation}}
\renewcommand\]{\end{equation}} 
\usepackage{listings}
\usepackage{fancyvrb}
\usepackage{framed}

\usepackage{courier}
\usepackage[usenames,dvipsnames,table]{xcolor}

\definecolor{dkgreen}{rgb}{0,0.3,0}
\definecolor{gray}{rgb}{0.5,0.5,0.5}

\makeatletter
\newcommand*{\rom}[1]{\expandafter\@slowromancap\romannumeral #1@}
\makeatother

\usepackage{siunitx}
\usepackage{tabu}
\usepackage{booktabs}
\usepackage{multirow}
\usepackage{capt-of}
\usepackage{array}
\usepackage{arydshln}
\usepackage{cite}

\newcommand{\comment}[1]{}

\makeatletter
\patchcmd{\@maketitle}
  {\addvspace{0.5\baselineskip}\egroup}
  {\addvspace{-1\baselineskip}\egroup}
  {}
  {}
\makeatother
\begin{document}

\title{




Intelligible Protocol Learning for Resource Allocation in 6G O-RAN Slicing

}

\author{
Farhad~Rezazadeh,~\IEEEmembership{Member,~IEEE}, 
Hatim~Chergui,~\IEEEmembership{Senior~Member,~IEEE}, Shuaib Siddiqui,
Josep~Mangues,
Houbing Song,~\IEEEmembership{Fellow,~IEEE},
Walid~Saad,~\IEEEmembership{Fellow,~IEEE}, and Mehdi~Bennis,~\IEEEmembership{Fellow,~IEEE}


\IEEEcompsocitemizethanks{\IEEEcompsocthanksitem Farhad Rezazadeh (corresponding author) is with the Telecommunications Technological Center of Catalonia (CTTC), Spain.}
\IEEEcompsocitemizethanks{\IEEEcompsocthanksitem Hatim Chergui and Shuaib Siddiqui are with the i2CAT Foundation, Spain.}
\IEEEcompsocitemizethanks{\IEEEcompsocthanksitem Josep Mangues is with the Telecommunications Technological Center of Catalonia (CTTC), Spain.}
\IEEEcompsocitemizethanks{\IEEEcompsocthanksitem Houbing Song is with the University of Maryland, Baltimore County (UMBC), USA.}
\IEEEcompsocitemizethanks{\IEEEcompsocthanksitem Walid Saad is with the Electrical and Computer Engineering Department, Virginia Tech, USA.}
\IEEEcompsocitemizethanks{\IEEEcompsocthanksitem Mehdi Bennis is with the University of Oulu, Finland.}}
\maketitle

\begin{abstract}
An adaptive standardized protocol is essential for addressing inter-slice resource contention and conflict in network slicing. Traditional protocol standardization is a cumbersome task that yields hardcoded predefined protocols, resulting in increased costs and delayed rollout. Going beyond these limitations, this paper proposes a novel multi-agent deep reinforcement learning (MADRL) communication framework called \emph{standalone explainable protocol (STEP)} for future sixth-generation (6G) open radio access network (O-RAN) slicing. As new conditions arise and affect network operation, resource orchestration agents adapt their communication messages to promote the emergence of a protocol on-the-fly, which enables the mitigation of conflict and resource contention between network slices.
STEP weaves together the notion of information bottleneck (IB) theory with deep Q-network (DQN) learning concepts. By incorporating a stochastic bottleneck layer---inspired by variational autoencoders (VAEs)---STEP imposes an information-theoretic constraint for emergent inter-agent communication. This ensures that agents exchange concise and meaningful information, preventing resource waste and enhancing the overall system performance. The learned protocols enhance interpretability, laying a robust foundation for standardizing next-generation 6G networks. By considering an O-RAN compliant network slicing resource allocation problem, a conflict resolution protocol is developed. In particular, the results demonstrate that, on average, STEP reduces inter-slice conflicts by up to $6.06 \times$ compared to a predefined protocol method. Furthermore, in comparison with an MADRL baseline, STEP achieves $1.4 \times$ and $3.5 \times$ lower resource underutilization and latency, respectively.

\end{abstract}


\section{Introduction}
\IEEEPARstart{6}{G} standardization is instrumental in ensuring seamless interoperability and widespread acceptance of mobile technologies such as network slicing. This technology relies on softwarization and cloudification approach that first emerged in the fifth-generation (5G) and is expected to gain more prominence in 6G. Indeed, 6G network slicing enables mobile network operators and service providers to deploy various virtual networks on a single physical infrastructure. The architectures of these network instances are fine-tuned for specific services such as ultra-reliable low latency communications (URLLC), enhanced mobile broadband (eMBB), and massive machine-type communications (mMTC). In this intent, resource allocation protocols are essential to maintain network service quality consistent in critical slices such as autonomous driving, unmanned aerial vehicles (UAVs), or remote surgery.  Given the broader range of applications and service-based architectures in 6G, opting for predefined standardized protocols for each architecture would increase system complexity, resulting in heightened costs, operational overhead, and a delayed rollout. Hence, more flexible protocols are needed to adapt to the plurality of 6G O-RAN slicing resource quota policies~\cite{o-ran-slicing}. This is essential to ensure consistent performance across slices with distinct bandwidth, latency and reliability requirements.

Artificial intelligence (AI) and machine learning (ML) approaches, particularly reinforcement learning (RL), can significantly address the challenges of standardized protocols in 6G network slicing. By viewing a protocol as a communicative language among 6G O-RAN network entities, intelligent network nodes can negotiate through message exchange \cite{Walid_paper_seman} to bring forth on-the-fly dynamic protocols tailored to the variety of concurrent slices and network contexts. Specifically, flexible resource allocation protocols are needed to address the inherent problem of inter-slice conflict and underutilization, and thereby ensure their isolation. This communication aspect would enable borrowing resources from less critical slices or reallocating bandwidth and computing resources from slices with surplus.

Indeed, over time and through interaction with the network slicing environment, a protocol can emerge among the agents in which specific communication messages correspond to particular meanings or behavioral cues. If agents discover that specific communication messages yield better outcomes, they will be more inclined to use them in subsequent interactions. This preference forms the basis for an \emph{emergent communication protocol}. A significant advantage of this emergent communication approach is adaptability. If the state of the network environment changes or agents identify better strategies, the communication protocol can evolve. Since the semantics and meanings of the communication codes are not hardcoded, agents have the flexibility to repurpose them as needed.

Recent works such as~\cite{9432398, Miuccio2022LearningGW, miuccio2023learning, Marwa_Chafii_2023} laid some of the groundwork for communication protocols, negotiation, adaptability, convergence time, and emergent communication. The authors in \cite{9432398} 
investigated whether radios can autonomously learn pre-established protocols and potentially evolve their own. Their findings illustrate that RL agents can be trained to learn basic medium access control (MAC) signaling and a wireless channel access strategy effectively. In~\cite{Miuccio2022LearningGW}, the authors 
unveiled an architecture combining autoencoders (AE) and a multi-agent proximal policy optimization (MAPPO) framework. Their findings underscore the potential of abstraction in protocol learning, showcasing enhanced robustness and generalization across varying environments. Correspondingly, the work in \cite{miuccio2023learning} 
proposed a reward system using a long short-term memory (LSTM) network that allows quicker training convergence in 6G wireless network MAC protocol design. 
Recently,
~the authors in ~\cite{Marwa_Chafii_2023} investigated the potential of emergent communication for upcoming 6G wireless networks.

Previous studies~\cite{9432398, Miuccio2022LearningGW, miuccio2023learning, Marwa_Chafii_2023} did not effectively confront the issues in resource allocation protocols towards 6G O-RAN network slicing.
The main contribution of this paper is a novel framework called ~\emph{standalone explainable protocol (STEP)}. STEP aims at learning an inter-slice protocol to minimize both computing resources underutilization and conflict. It is an additional building block that fosters collaboration, reduces competition between the agents of concurrent slices in AI-native 6G networks, and does not interfere with or replace the existing network protocol stack.

It is powered by the information bottleneck (IB)~\cite{Shao-IB, Geiger} framework to extract the most relevant information from slice state and received messages, thereby yielding efficient communication and decision-making. We propose a stochastic bottleneck layer to prevent agents from getting stuck in deterministic patterns, allowing for adaptability to unforeseen network requirements. The stochastic bottleneck is a neural network layer designed to introduce stochasticity into the model. This is particularly important for conflict resolution in network slicing, where deterministic patterns might lead to repeated conflicts and inefficient resource utilization. The stochastic bottleneck employs the reparameterization technique to allow backpropagation through random nodes. Integrating IB theory synergistically with stochastic bottleneck layer within our proposed STEP framework can pave the way for AI models with optimized information flow. This enhancement can lead to more interpretable models~\cite{Roy-XAI} that compel the resource orchestration agents to focus on the most pertinent and succinct information.

To delve into the abovementioned aspects, in Section~\ref{sec:chal_limit}, we discuss some key technical challenges of network slicing. In Section~\ref{sec:framework_step}, we explain how various components of the STEP frameworks interplay. Moreover, we outline the compliance of STEP with O-RAN slicing. Then, the efficacy and performance metrics of the STEP-based resource allocation are showcased in Section~\ref{sec:eval_results}. Section~\ref{sec:fut_research} sheds light on future research directions in the context of protocol learning in network slicing.

\section{Technical Challenges in 6G Network Slicing}
\label{sec:chal_limit}
\subsection{Agent Definition}
\label{sec:chal_limit_agentdef}
Network slicing leverages virtual network functions (VNFs)~\cite{vnf-Motalleb} or container-based NFs (CNFs), each configured/placed to meet specific requirements of a slice (service). These VNFs/CNFs-based isolated slices span geo-distributed nodes and do not always coincide in the same node, but they correlate since they are on the same cluster. Given the non-scalability of a centralized orchestrator (usually located in a separate cluster) due to the associated overhead, longer reaction time, and single point of failure, distributed slice orchestration agents are needed. Here, the agents are software entities (DRL-based) responsible for autonomous decision-making tasks that need to adapt their communication methods as network conditions change. This emergent communication is vital in network slicing dynamic environments, as their software code and algorithms drive the decision-making logic.
\subsection{Inter-slice Decisions Conflicts}
\label{sec:Inter_slice_limit}

In O-RAN network slicing, efficient statistical multiplexing among slices is highly desirable due to infrastructure costs and the dynamic traffic loads of each service \cite{spatial}. Hence, although a slice is initially allocated a share of a pool of distributed resources, they can be preempted by its concurrent slices, ideally, if they are unused. In absence of coordination, the agents orchestrating the different parallel slices might decide to allocate more resources than their shares in a simultaneous way---based on their local observation space, which would lead to an inter-slice decision conflict. Therefore, the key challenge is managing resource allocation among concurrent slices to avoid inter-slice conflicts.
\subsection{Isolation vs. Monitoring Data Sharing}
\label{sec:monit_chall}
The fundamental idea of maintaining isolation between network slices is a key aspect of 6G networks. One facet of this isolation involves monitoring, ensuring that the measured raw performance data of a particular slice remains separate from another slice that shares the same infrastructure. This is necessary for security but also for scalability considerations, since extensive inter-slice monitoring data sharing can strain the network and incur additional overhead. Given that centralized slice orchestration is not scalable, the adoption of distributed decision agents is required. This entails exchanging monitoring information between slices for the purpose of improving the decision-making, which could violate the notion of isolation. Alternatively, agents need to share an abstracted state that does not reveal the raw data.

\subsection{Poor Generalization in Unseen Tasks}
\label{sec:gen_tasks}
As mentioned in Subsection \ref{sec:monit_chall}, network slices' resources are orchestrated by agents to achieve the local objective of each slice, such as minimizing latency and/or maximizing throughput. This optimization process is accomplished by continually observing the network states and making informed decisions, such as allocating bandwidth or computing resources. These deep reinforcement learning (DRL) agents leverage deep neural networks (DNNs) to estimate policy or value functions. Nonetheless, DNNs have a tendency to memorize specific environmental observations rather than understand the underlying patterns or dynamics, which limits their ability to generalize to environments and tasks unseen during training, such as different traffic behaviors. Such challenges become even more pronounced in network slicing, where multiple agents interact and influence one another.
\section{STEP Framework}
\label{sec:framework_step}

We propose to address the challenges outlined in Section~\ref{sec:chal_limit} by developing STEP---a novel framework that combines multi-agent DRL with the IB theory in order to ensures that the transmitted data 
undergoes a principled reduction, preserving only the most relevant information while discarding the redundant and superfluous information. 


\begin{figure}[t!]
\centering
\includegraphics[scale=0.65]{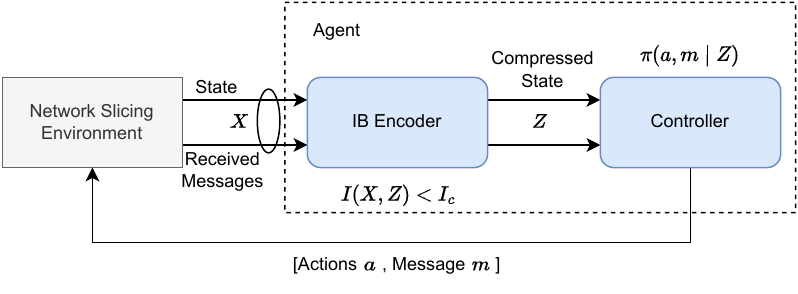}
\caption{\small The STEP workflow for an individual agent involves observing a compressed version of state and communication spaces, thereby discarding redundancy and facilitating the interpretation of emerging protocols. The network slicing environment is detailed in Fig.~\ref{fig:XDRL_architecture}.}
\label{fig:STEP_architecture}
\vspace{-5mm}
\end{figure}
\subsection{Intelligible Protocol Learning}
\label{sec:Emergent_comm}

Within 6G O-RAN environments, resource orchestration agents might cooperate through inter-agent communication. In this regard, they can autonomously develop a communication protocol without being explicitly taught.
Fig.~\ref{fig:STEP_architecture} demonstrates the generic workflow of a single DRL agent in STEP approach. STEP seeks to solve the discussed challenges in Section~\ref{sec:chal_limit}. For instance, consider a gNodeB (gNB) slicing environment with three different slices A (URLLC), B (eMBB), and C (mMTC), where each slice is equipped with a STEP agent. Existing MADRL methods in network slicing require the slices to be informed about other slices' network states (i.e., monitoring information) and focus solely on resource allocation actions. In contrast, the proposed STEP agent operates differently in that it  does not need this monitoring information. Instead, STEP agents are designed to learn dual-actions, which consist on resource allocation (as a primary action) and control messages (as a secondary one). This method addresses the challenges outlined in Section~\ref{sec:monit_chall}. Specifically, agent A begins by observing the states of the network slice A, and it receives signals from agents B and C. Messages conveyed between agents lack predetermined significance. The agents develop \emph{communication strategies on-the-fly} based on the network environment's feedback. The observations are processed as an input vector, which feeds a DNN. Within this network, the information undergoes a sequence of transformations to cope with the challenges in Section~\ref{sec:gen_tasks}. 

The input is initially processed by a fully connected layer, converting it into an intermediate representation. Then, the framework leverages a \emph{stochastic information bottleneck} within the neural architecture. Overfitting is a recognized problem in DRL where a model is good at handling familiar scenarios but struggles with unseen situations. The model might memorize particular environmental states and their corresponding rewards rather than understanding the broader dynamics or patterns. To solve this, we define $I(X,Z)$ as the mutual information between compressed state $Z$ and state $X$, i.e., how much information $Z$ provides about the original state $X$ concerning constraint $I_c$. This bottleneck processes information by retaining only the most essential features, which are then used to determine Q-values for decision-making in the controller. In this regard, a Kullback-Leibler (KL) divergence
~term is added to the loss function to ensure the bottleneck's output remains relevant and retains only the most crucial information while discarding the rest, acting as a form of regularization. This approach ensures the output aligns with a specific prior distribution. By adding a randomized component (sample from a standard normal distribution), the stochastic bottleneck enables agents to understand how they explore different network states and messages with a degree of uncertainty. This reparameterization trick promotes exploration and assists agents in evolving more effective communication strategies over time. Indeed, stochastic information bottleneck forces the network to focus on broader patterns and dynamics rather than specific states. After passing through the bottleneck, the processed information is directed to a subsequent fully connected layer. This layer produces Q-values, representing the agent's estimates of the benefits of specific actions (e.g., allocating radio or computing resources to the slice). The agent then selects an action based on the Q-values. The total loss combines the Q-learning loss and the KL divergence loss, modulated by a coefficient back-propagated to the entire network to update the parameters. Section~\ref{sec:Evaluation-results} delves deeper into evaluation metrics and discusses how the STEP framework solves the problems in Section~\ref{sec:Inter_slice_limit}.

\subsection{O-RAN Slicing Compliance}
\label{sec:o-ran}
Our STEP framework can align with the objectives of O-RAN slicing use case $three$~\cite{o-ran-slicing}. This use case is centered on improving resource allocation for network slice subnet instances (NSSIs) in 6G networks. This involves the implementation of resource quota policies designed to enhance resource utilization across NSSIs. These policies also provide a basis for allocating resources according to service priority and historical usage data. In the O-RAN ecosystem, the policy outcomes trained by STEP can be instantiated as rApps within the non-real-time (Non-RT) radio intelligent controller (RIC). The STEP's underpinning architecture enhances this integration, synergizing explainability with continuous integration and continuous deployment (CI/CD), granting the system exceptional agility and automation capabilities. To ensure these rApps execute their functions precisely, they access detailed monitoring data from the O1 interface. Once the STEP model undergoes rigorous training, it is packaged as a deployable artifact. This artifact is subsequently relayed via the A1 interface, allowing it to operate on the Near-RT RIC interface as an xApp. REST APIs support this deployment. The main goal of this xApp is dynamic network optimization, which governs the underlying O-RAN technological domains using the E2 interface. This interface is a composite of two key protocols, the E2 application protocol (E2AP) and the E2 service model (E2SM), ensuring efficient and standardized communication within the O-RAN framework. Specifically, STEP can enhance the efficiency of O-RAN CPU quota policies among various network slices \cite{o-ran-slicing}. Such a capability is especially beneficial in prioritizing critical service slices or optimally reallocating CPU resources in response to changing network conditions.

\section{Performance Evaluation of STEP Framework}
\label{sec:eval_results}
Without loss of generality, we exemplify how the STEP framework can lead to the emergence of efficient inter-slice conflict and underutilization resolution protocols. We validate the STEP's efficacy, focusing on improvements in latency, CPU utilization, and reducing the number of conflicts. Moreover, we analyze agents' performance under various communication message sizes and explainability metrics.

\subsection{Use Case: Learning Inter-Slice Conflict Resolution Protocol}

\begin{figure}[t!]
\centering
\includegraphics[scale=0.15]{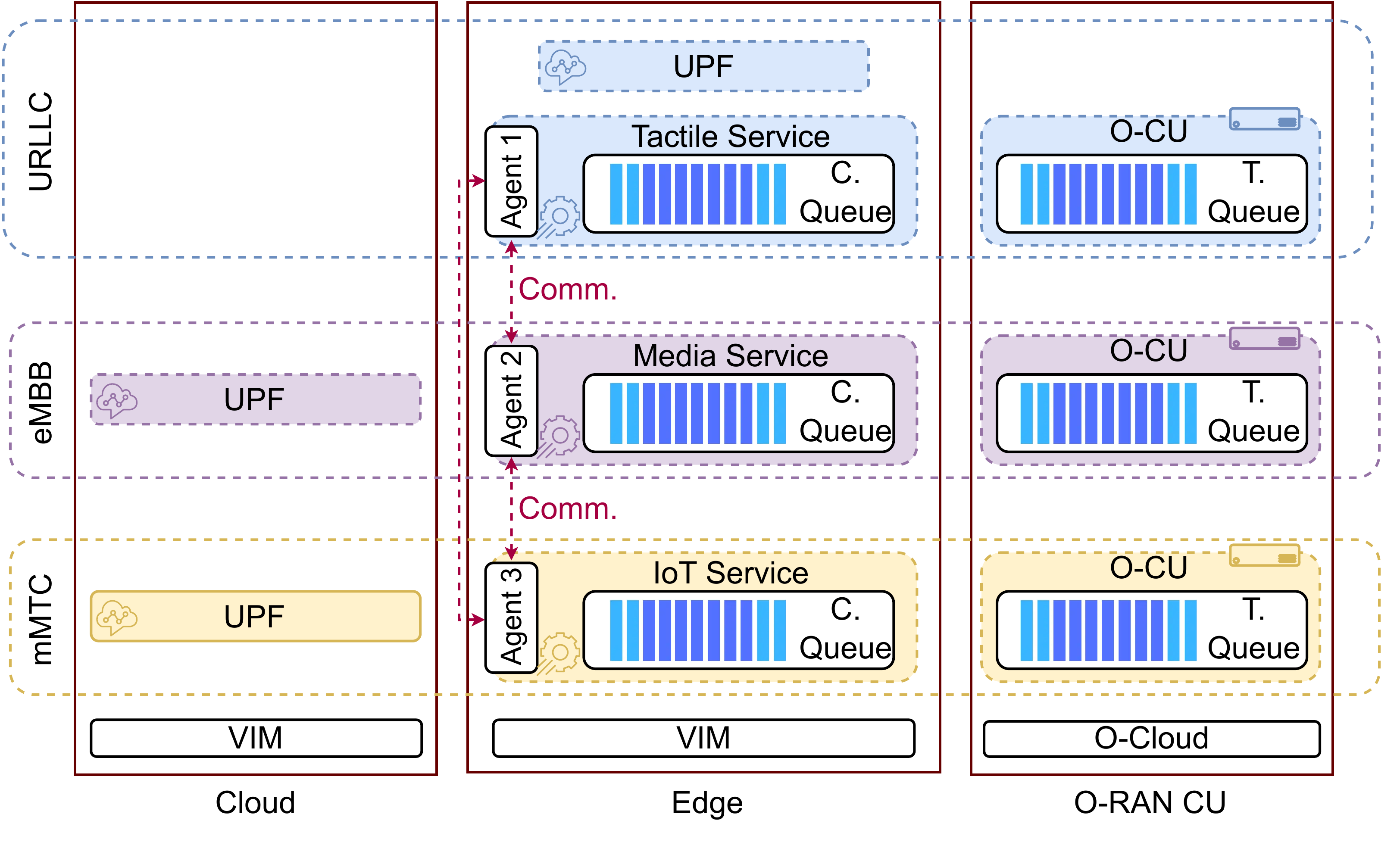}
\caption{\small Architecture of the inter-slice conflict resolution use case, with one agent per slice, where the O-RAN and Edge domains form the network slicing environment.}
\label{fig:XDRL_architecture}
\end{figure}

Experimental evaluations have demonstrated a non-linear relationship between bitrate, virtualized RAN (vRAN) bandwidth, and CPU utilization. Even with sufficient radio resources, network performance suffers without adequate computational resources in the O-Cloud. Thus, strategic and coordinated CPU allocation is imperative for effective radio resource management (RRM) in O-RAN's vRAN instantiation~\cite{Hervas-o-cloud, Kubernetes2024ResourceQuota, Moroclould}. As shown in Fig. \ref{fig:XDRL_architecture}, an inter-slice intelligent resource orchestration use-case is considered, where each slice is composed of a server located at the edge domain. The virtual infrastructure manager (VIM) manages the computing, storage, and network resources in a virtualized infrastructure, typically in a cloud environment. It has a computation queue (C. queue) with preemptive CPU resources whose dynamics impact the local latency. The O-RAN domain is endowed with a per-slice transmission queue (T. queue) at the O-CU level, whose latency is affected by the radio conditions. By omitting the x-haul delays, the slice latency is the aggregation of the O-RAN and edge ones. As a physical action, each slice-level resource orchestration agent scales the CPU frequency and can use more than its resource share if the concurrent slices leave underutilized resources. Given the non-scalability of centralized approaches, the agents (see Section~\ref{sec:chal_limit_agentdef}) collaborate through a decentralized multi-agent deep reinforcement learning framework, where they coordinate via a so-called communication action or message. The DRL agents must cooperatively learn/discover the signaling policy, even without prior agreement on the meaning of control messages. They are guided by a reward function that penalizes conflicts and underutilization and minimizes latency. By properly scaling the CPU frequency the agent minimizes latency, while the messages allow for avoiding conflict and underutilization.

\subsection{Network Architecture and Experiment}
\label{server-spec}
We consider a scenario with service-specific slices. In this context, slices are represented by data sourced from a simulator \cite{Specialization_TVT}, which captures the nuances of services like eMBB, URLLC, and mMTC in terms of traffic and corresponding radio resources, while the queues are developed as a Python add-on. The simulator comprises three core components: physical (PHY), MAC, and radio link control (RLC) functions. It also includes the O-RAN E2 interface, which is used for collecting network statistics from each distributed unit (O-DU). To avoid underutilization, each slice has a share of CPU resources that can also be used by other slices as long as this does not lead to a conflict. Each slice is equipped with a STEP agent. The agents aim to optimize the allocation of CPU frequencies to slices and exchange messages to solve inter-slice resource contention and conflict by fostering the emergence of a conflict resolution protocol. The CPU resource allocation is defined by the array [15, 15, 10], representing the default CPU shares for each of the three slices in $10^9$ cycles per second units. This configuration means that the first and second slices are each allocated 15 GHz, and the third is allocated 10 GHz under default conditions. This setup assumes that the cores can efficiently handle parallel tasks. Note that we can apply the same approach to other resource allocation problems, such as physical resource blocks (PRB) allocation problem. Each agent has a DQN incorporating an IB block. As we discussed in Section~\ref{sec:Emergent_comm}, the IB, with a \textit{stochastic bottleneck} layer, is designed to capture essential features from the input while discarding redundant information. 
This approach is assisted by a prioritized experience replay to sample experiences non-uniformly based on their importance, focus on the most pivotal network states, and ensure generalizing across various scenarios.

To ensure the generalizability of STEP, we conducted a \emph{time-series} and \emph{variability} analysis on the traffic across three slices. This approach offers valuable insights into STEP's adaptability to varying traffic conditions, which are quantified by different standard deviations across various time periods. Specifically, for eMBB Traffic, the mean is 23.33 Mbps with a standard deviation of 22.38 Mbps. For URLLC Traffic, the mean is 7.80 Mbps with a standard deviation of 9.25 Mbps. Lastly, for mMTC Traffic, the mean is 14.80 Mbps, accompanied by a standard deviation of 20.86 Mbps.

The designed network slicing environment in Fig.~\ref{fig:XDRL_architecture} simulates the interactions among agents and their impact on the overall network. As agents optimize their resource allocations (CPU frequencies) and send message to other agents, the environment updates the system state (i.e., served traffic, cpu allocation gap, and received messages by other agents) and returns the corresponding rewards. The rewards are designed to penalize conflicts and heavily reward low latency. Thus, the final reward is a sum of both of these metrics. For instance, the main component of the reward is an exponential function of the negative latency to ensure when network latency decreases; the reward will be increased. However, the agents are penalized heavily when resource constraints are violated, i.e., the total allocated CPU frequencies exceed the maximum available. During training, the agents employ an \(\epsilon\)-greedy strategy for exploration and exploitation, with \(\epsilon\) decaying over episodes. 
Indeed, as agents decide on resource allocations and exchange messages, the environment returns feedback regarding system states and rewards. 
The agents then use this feedback to adapt and optimize resource allocation strategies. They strive for minimized latencies by allocating optimal CPU frequencies while avoiding inter-slice resource conflicts. 

Our experiments are conducted on a dedicated server featuring two Intel(R) Xeon(R) Gold 5218 CPUs @ 2.30GHz, dual NVIDIA GeForce RTX 2080 Ti GPUs. The DNN network comprises an input layer mapped to a 64-neuron hidden layer activated by the ReLU function. The subsequent IB layer, designed for feature compression, employs a reparameterization trick with a bottleneck dimension 32. Unlike simple compression techniques that just aim to reduce data size or complexity, our approach provides a trade-off between compression and preservation of relevant information. This trade-off is controlled by a parameter $\beta$, which is incorporated into the loss function. Indeed, $\beta$ is a regularization parameter that balances the two components of the total loss: the primary Q-learning loss and the IB loss. A higher $\beta$ places more emphasis on the information bottleneck constraint, leading to more compression (or information reduction) at the cost of potentially losing some relevant information.
Finally, the network uses a fully connected layer outputting both main actions and messages. During training, weights are updated using the Adam optimizer with a learning rate of $0.0005$. The training process is executed for a total of \(500\) episodes with a maximum of \(60\) steps per episode. The agents are trained with a batch size of \(64\), a discount factor (\(\gamma\)) of \(0.99\), and an initial \(\beta\) of \(0.0\) that anneals at a rate of \(0.001\) per episode. This simulation, built using libraries like \texttt{torch} and \texttt{gym}, is an essential step towards validation of resource allocation in a network slicing scenario and provides insights into optimizing service-specific slices.

\subsection{Numerical Results}
\label{sec:Evaluation-results}
\begin{figure}[t!]
\centering
\includegraphics[scale=0.7]{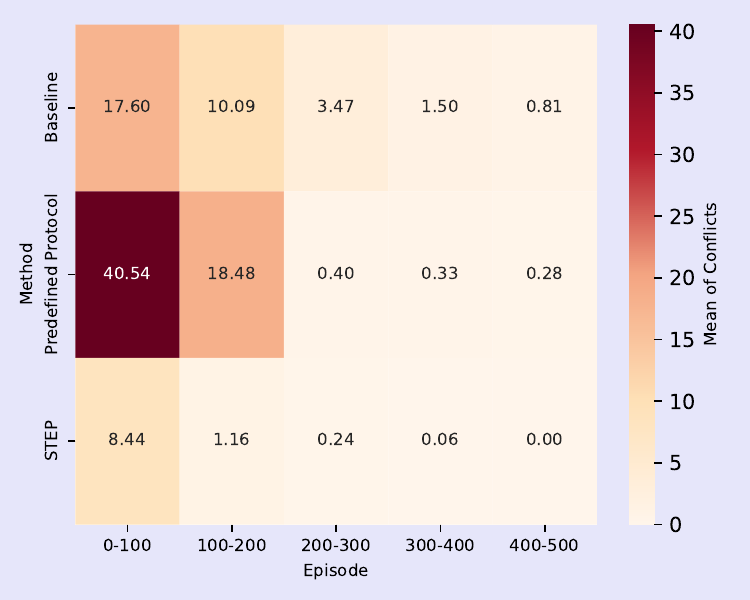}
\caption{\small Evaluation of average computing resource allocation conflict number among three agents. All assessments were carried out on the server delineated in Section~\ref{server-spec}.}
\label{fig:heatmap_approaches}
\end{figure}
\begin{figure}[t!]
\centering
\includegraphics[scale=0.7]{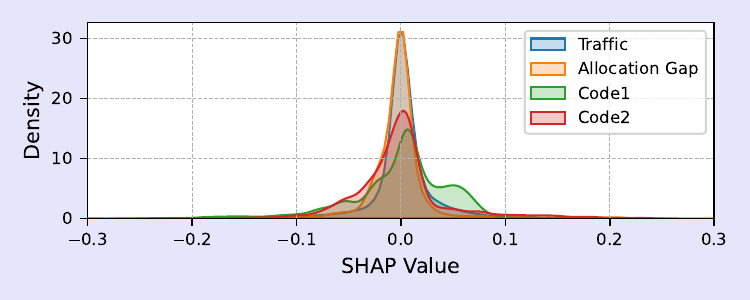}
\caption{\small The KDE plot represents the attribution of each feature to the model's predictions. The x-axis denotes the SHAP values, capturing the contribution of features to the prediction. The y-axis represents the density, indicating the frequency of occurrence of each SHAP value.
 }
\label{fig: msg-shap}
\end{figure}

Fig.~\ref{fig:heatmap_approaches} shows the average occurrences of resource conflicts between slices over successive 100-episode intervals during training. The figure clearly shows that in the first 100 episodes, the standard MADRL framework baseline, which does not incorporate emergent communication, averaged $17.6$ conflicts. In contrast, the STEP approach demonstrated a considerably lower average of $8.44$ conflicts. Alongside these methods, a predefined protocol approach was utilized, in which DRL agents communicated their resource allocation strategies through three distinct codes: 0 (indicating usage above their share), 1 (usage below their share), and 2 (exact usage of their share). In the STEP framework, the agents use these codes but may attribute to them different meanings which evolve according to the context. Using SHAP values, Fig. \ref{fig: msg-shap} shows the influence of each code on the resource allocation. However, this approach initially showed limited effectiveness, with an average of 40.54 conflicts observed. This demonstrates the enhanced efficiency of the STEP strategy from the beginning of training. As training progresses, the difference between strategies becomes evident. By the concluding set of $100$ episodes, the baseline strategy still encountered an average of 0.81 conflicts, while the STEP strategy impressively brought the conflict count down to zero. With its simple communication scheme, the predefined protocols method encountered an average of only 0.28 conflicts, indicating a significant improvement over the baseline, though not quite reaching the effectiveness of the STEP strategy. Specifically, on average, STEP reduces inter-slice conflicts by a factor of $3.4 \times$ and $6.06 \times$ compared to the baseline and predefined protocol methods, respectively. Clearly, the agents employing the STEP strategy learned an efficient protocol for resource allocation throughout training, eradicating conflicts. The STEP strategy's ability to adaptively learn between agents resulted in optimized conflict resolution, eliminating the need for a predefined resource allocation protocol.

\begin{figure}[t!]
\centering
\includegraphics[scale=0.7]{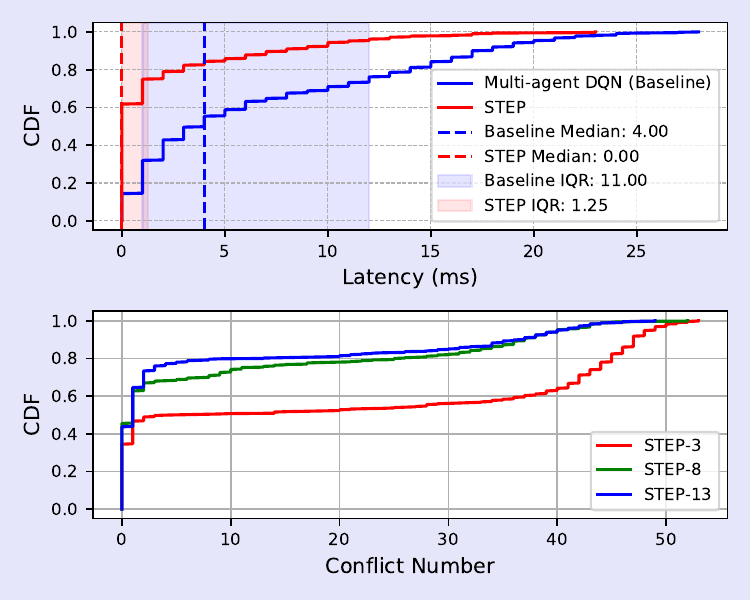}
\caption{\small (Up) CDF representation of average transmission and computation latencies across three slices, (Down) Comparative analysis of STEP performance regarding average resource allocation conflicts across three agents with varied communication action sizes. Unlike other experiments, this test was conducted under an exceptionally stringent conflict threshold.}
\label{fig:base-step-latency}
\end{figure}

In Fig.~\ref{fig:base-step-latency} (Up) the proposed STEP outperforms the baseline both in terms of average latency and consistency of performance. The CDF plot detailing the average latency across three slices shows the baseline's median latency at 4 ms. This signifies that half ($50\%$) of the recorded latencies are below this mark, while the other half exceed it. In contrast, the proposed STEP framework has a median latency of $0$ ms, indicating that half of its latencies are at $0$ ms. Notice that we omitted the x-haul delays and therefore the incurred slice latency is the aggregation of the O-RAN and edge queues' ones. This remarkable improvement over the baseline suggests that STEP processes traffic more swiftly on average than the baseline. The interquartile range (IQR) gives an insight into the spread or dispersion of the middle $50\%$ of the data. A smaller IQR indicates that the middle $50\%$ of the data points are densely clustered, which implies reduced variability. The IQR for the baseline is 11 ms, showing a broad spread in its latency values. Conversely, STEP's IQR stands at a mere $1.25$ ms, underscoring that its central $50\%$ of latency data is more consistent and exhibits minimal variability. On average, this results in $3.5\times$ lower latency compared to MADRL. 

The bottom of Fig.~\ref{fig:base-step-latency} investigated the conflict CDF for various communication space sizes, i.e., a set of messages or signals an agent sends or receives with other agents. While increasing the size offers more diversity in the exchanged messages, the resulting information flow might hinder the benefits of inter-agent communication. Indeed, we have the option to either abstain from exchanging messages altogether or engage in widespread message sharing. When no message is shared, we might revert to a purely competitive strategy, where a high number of conflicts occurs. On the other hand, when communication occurs with higher message alphabets, we risk diminishing diversity in our problem-solving approaches, potentially resulting in a homogeneous strategy where each slice essentially handles the same tasks, which justifies the quasi-similarity observed for $8$ and $13$. Therefore, there is a \emph{saturation point} for the communication space size. In Fig.~\ref{fig:base-step-latency}, we conducted both tests with a very strict limit on the network's computing resources. This was done to thoroughly test how well the STEP algorithm works in harsh scenarios.


\begin{figure}[t!]
\centering
\includegraphics[scale=0.7]{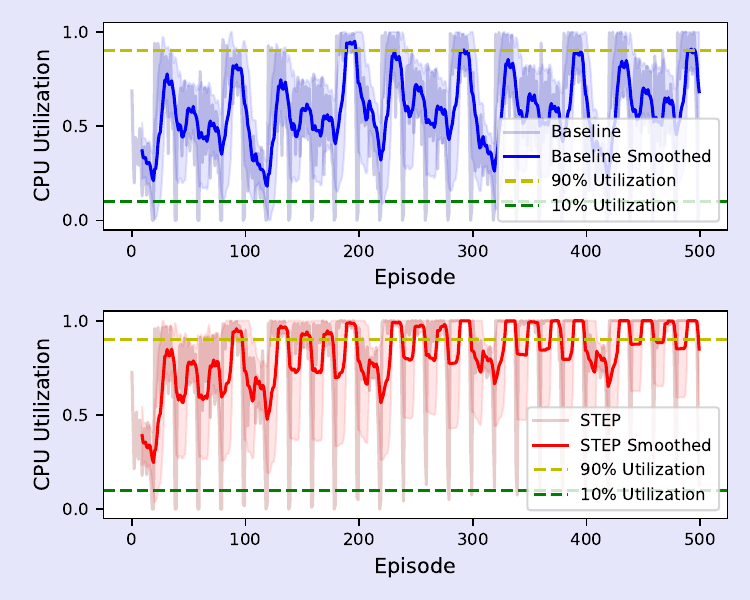}
\caption{\small A comparative view of baseline vs. STEP approaches in terms of CPU utilization performance. To provide a clearer visual representation, the curves are smoothed with respect to confidence bands and standard deviation.}
\label{fig:cpu_cdf}
\end{figure}

Fig.~\ref{fig:cpu_cdf} shows the CPU utilization of the MADRL baseline and the proposed STEP framework. By using the proposed emergent protocol, agents are able to detect free CPU resources of the rest of slices and use them temporarily to further boost their computation queue CPU frequency. This leads to higher levels of CPU network-wide utilization that surpass the lower and upper thresholds of $10\%$ and $90\%$, respectively, outperforming the baseline scheme, which struggles to keep a consistent utilization across episodes. Specifically, STEP approach reduces resource underutilization by up to a $1.4\times$ compared to the baseline.

Fig.~\ref{fig: msg-shap} shows the distribution of SHAP importance values for the different input observations and received messages. In Section~\ref{server-spec}, we discussed the \emph{traffic} in the context of network slicing, which pertains to the flow of data within distinct network slices. These slices are optimized for various service types, such as eMBB, URLLC, and mMTC. Additionally, the \emph{allocation gap} is defined as the difference between the resources allocated to each network slice and its actual required resource. Moreover, \emph{Code 1} and \emph{Code 2} represent the messages received by Agent 3 from Agent 1 and Agent 2, respectively. The traffic and allocation gap observations present a symmetric Gaussian behavior, where their variations can slightly lead to a scale-up or scale-down of CPU resources depending on the dominant objective (conflict reduction or latency minimization). Interestingly, the messages are more inclined to positively impact the CPU scaling with higher SHAP values range, which translates into higher utilization, while also exhibiting a more potent negative effect when it comes to conflict reduction.

\section{Future Research Directions}
\label{sec:fut_research}
\subsection{Communication Space Design} 
Studying the effect of the communication space design, scheduling, and channel conditions on the learning performance is an inter-sectoral research topic, where both AI and information theoretic tools can be leveraged to derive channel-aware closed-forms or information theoretic bounds on the AI performance. Additionally, the choice of the right inter-agent communication channel, such as cheap-talk, is an open research direction.

\subsection{Scalability of Protocol Learning}
With the growth in the number of agents and intricacy of their interactions, ensuring scalability in protocol learning becomes difficult. 
The meta-learning approach can enables agents to rapidly adapt to new protocols using their past experiences. This might enhance scalability and enable agents to fine-tune their protocols to adapt to new situations rapidly.
\subsection{Cross-Layer Protocols}
Conventionally, communication protocols are designed to operate within specific layers of the communication stack. This layered approach may make decisions without considering the other layers' states, resulting in suboptimal resource allocation decisions. The cross-layer protocol learning is an open research line, which consists of learning protocols that span multiple layers of the communication stack to more comprehensively optimize resource overheads and ensure seamless integration with prevailing standards.
\section{Conclusion}
In this paper, we have proposed a novel STEP framework that exploits multi-agent DRL combined with IB theory to enable agents to adapt and optimize their communication protocols on-the-fly. The results show that agents trained within STEP approach can create efficient communication protocols 
by imposing information-theoretic constraints, compressing thereby the state information and leading to resource efficiency. 
We have then developed an example that uses STEP to develop an inter-slice conflict and underutilization resolution protocol. Our results have shown the advantage of STEP in terms of conflict and latency reduction, and utilization improvement compared to the MADRL baseline. 

\section*{Acknowledgment}
This work was partially funded by MCIN/AEI/ 10.13039/501100011033 grant PID2021-126431OB-I00 (ANEMONE), Spanish MINECO grant TSI-063000-2021-54 (6G-DAWN) and grant TSI-063000-2021-56 (6G-BLUR), Generalitat de Catalunya grant 2021 SGR 00770 (6GE2E), the Horizon Europe projects NANCY (101096456) and COGNIFOG (101092968), the European Union through projects CENTRIC (G.A no. 101096379), the U.S. National Science Foundation through Grant No. 2317117, Grant No. 2309760, and Grant No. CNS-2225511,


\section*{Biographies}
\noindent Farhad Rezazadeh [S'19, M'23] (farhad.rh@ieee.org) received the Ph.D. degree (Excellent Cum Laude) in Signal Theory and Communications from Technical University of Catalonia (UPC), Spain. He is currently an R\&D engineer at the CTTC, Spain. He participated in 8 European and National 5G/B5G/6G R\&D projects with leading and technical tasks in the areas of Applied AI. He was a secondee at NEC Lab Europe and had scientific missions at TUM, Germany, TUHH, Germany and UdG, Spain. He is a Marie Sklodowska-Curie Ph.D. grantee and won 5 different IEEE grants, 2 COST grants, and a Catalan Government Ph.D. Grant. He actively serves as Organizing, Chair, Reviewer, and TPC member in IEEE and Guest Editor in Elsevier.  
\\
\\
\noindent Hatim Chergui [SM'22] (chergui@ieee.org) is a Senior Researcher at i2CAT Foundation, Spain. He was the project manager of the H2020 MonB5G European project and a researcher at CTTC, Spain. He served as a RAN expert at both INWI and Huawei Technologies, Morocco. He has published more than 40 research papers in top-tier journals and conferences and has contributed to 1 European patent. He was the recipient of the IEEE ComSoc CSIM 2021 Best Journal Paper Award and the IEEE ICC 2020 Best Paper Award. He is an Associate Editor of IEEE Networking Letters and has been a Chair in several IEE Workshops.
\\
\\
\noindent Shuaib Siddiqui (shuaib.siddiqui@i2cat.net) received the Ph.D. degree in computer science from the Technical University of Catalonia, Spain. He is a Senior Researcher with i2CAT Foundation where he is also the Area Manager with Software Networks research Lab. He is also currently coordinating the H2020 5GZORRO (5G PPP Phase 3 Project). His current research topics include network automation, SDN/NFV based control, management, and orchestration platforms for 5G, network slicing, and NFV/SDN security.
\\
\\
\noindent Josep Mangues~(josep.mangues@cttc.cat)~received the PhD degree in Telecommunications in 2003 from the Technical University of Catalonia (UPC). He is Research Director and Head of Services as networkS research unit of the CTTC. He has published 120+ journals/magazines and international conference papers through collaborations with multiple research groups. Since July 1997 he has participated in around 40 EU, Spanish, and industrial research projects in various roles (incl. PI). Interests include Cloud/Edge computing, NFV in mobile networks (incl. open RAN), data science/engineering, AI/ML-based service and network management automation and orchestration.
\\
\\
\noindent Houbing Song~[F]~(songh@umbc.edu) received the Ph.D. degree in electrical engineering from the University of Virginia, Charlottesville, VA, USA, in August 2012. He has been serving as an Associate Technical Editor, Associate Editor, and Guest Editor for IEEE journals and Transactions. He is an ACM Distinguished Member and an ACM Distinguished Speaker. He was a recipient of more than best paper awards from major international conferences, including the IEEE CPSCom-2019, the IEEE ICII 2019, the IEEE/AIAA ICNS 2019, the IEEE CBDCom 2020, the WASA 2020, the AIAA/IEEE DASC 2021, the IEEE GLOBECOM 2021, and the IEEE INFOCOM 2022.
\\
\\
\noindent Walid Saad [S'07, M'10, SM'15, F'19]~(walids@vt.edu) received his Ph.D degree from the University of Oslo, Norway in 2010. He is currently a Professor at the Department of Electrical and Computer Engineering at Virginia Tech, where he leads the Network sciEnce, Wireless, and Security (NEWS) laboratory. His research interests include wireless networks (5G/6G/beyond), machine learning, game theory, security, UAVs, semantic communications, cyber-physical systems, and network science. He is also the recipient of the NSF CAREER award in 2013, the AFOSR summer faculty fellowship in 2014. He is the Editor-in-Chief for the IEEE Transactions on Machine Learning in Communications and Networking.
\\
\\
\noindent Mehdi Bennis~[F]~(mehdi.bennis@oulu.fi) is currently a tenured Full Professor with the Centre for Wireless Communications, University of Oulu, Finland,
and the Head of the Intelligent COnnectivity and Networks/Systems Group (ICON). He has published more than 200 research papers in international conferences, journals, and book chapters. His research interests include radio resource management, game theory, and distributed AI in 5G/6G
networks. He has been a recipient of several prestigious awards. He is an Editor of IEEE TRANSACTIONS ON COMMUNICATIONS and a specialty Chief Editor of Data Science for Communications and Frontiers in Communications and Networks journal.
\end{document}